\documentclass{article}
\usepackage[utf8]{inputenc}
\usepackage{graphicx}
\usepackage{caption}
\usepackage[subrefformat=parens]{subcaption}
\usepackage{authblk}

\title{
Simulation results of a New type of sandwich calorimeter, Double readout Sandwich Calorimeter (DSC) performance
}

\author[1]{T.~Takeshita}
\author[1]{R.~Terada}

\affil[1]{Department of Physics, Faculty of Science, Shinshu University}

\begin{document}

\maketitle

\vspace{10pt}

\abstract{
In this study, we propose a novel fully active total absorption
calorimeter with enhanced readout capabilities through fine splitting.
The proposed calorimeter employs a sandwich structure composed of scintillator glasses and lead glasses, which are subdivided into small tiles.
This configuration enables the creation of a finely segmented three-dimensional calorimeter that is compatible with Particle Flow Algorithms (PFA).
This article presents a comprehensive description of the calorimeter design and reports performance results obtained through simulations.
Notable, calorimeter demonstrates exceptional linearity and energy resolution, closely rivaling that of homogeneous calorimeters.
The simulated calorimeter achieves an impressive energy resolution of approximately $ 9\% / \sqrt{E(\textrm{GeV})}$.
}

\newpage

\section{Introduction}
Future high-energy experiments in particle physics necessitate substantial advancements in the energy resolution of hadron calorimeters. 
While homogeneous calorimeters constructed from a single material are known for their excellent energy resolution, they face challenges related to light transmission, radiation-tolerance, segmentation, and cost \cite{bib:1}. 
In this study, we propose a novel approach to address these challenges by introducing a fully active, three-dimensional segmented calorimeter employing two similar materials: scintillator glass and lead glass which has cost advantage. By combining these materials in a sandwich structure, we aim to achieve a finely segmented calorimeter that maintains high energy resolution while mitigating the limitations associated with large homogeneous calorimeters.
While these materials bear few distinctions, the scintillator glass measures energy through scintillation light, whereas the lead glass, excels in detecting Cerenkov light, which is directory proportional to the length of charged particle trajectories.

\begin{figure}[ht!]
  \centering
   \includegraphics[width=80mm]{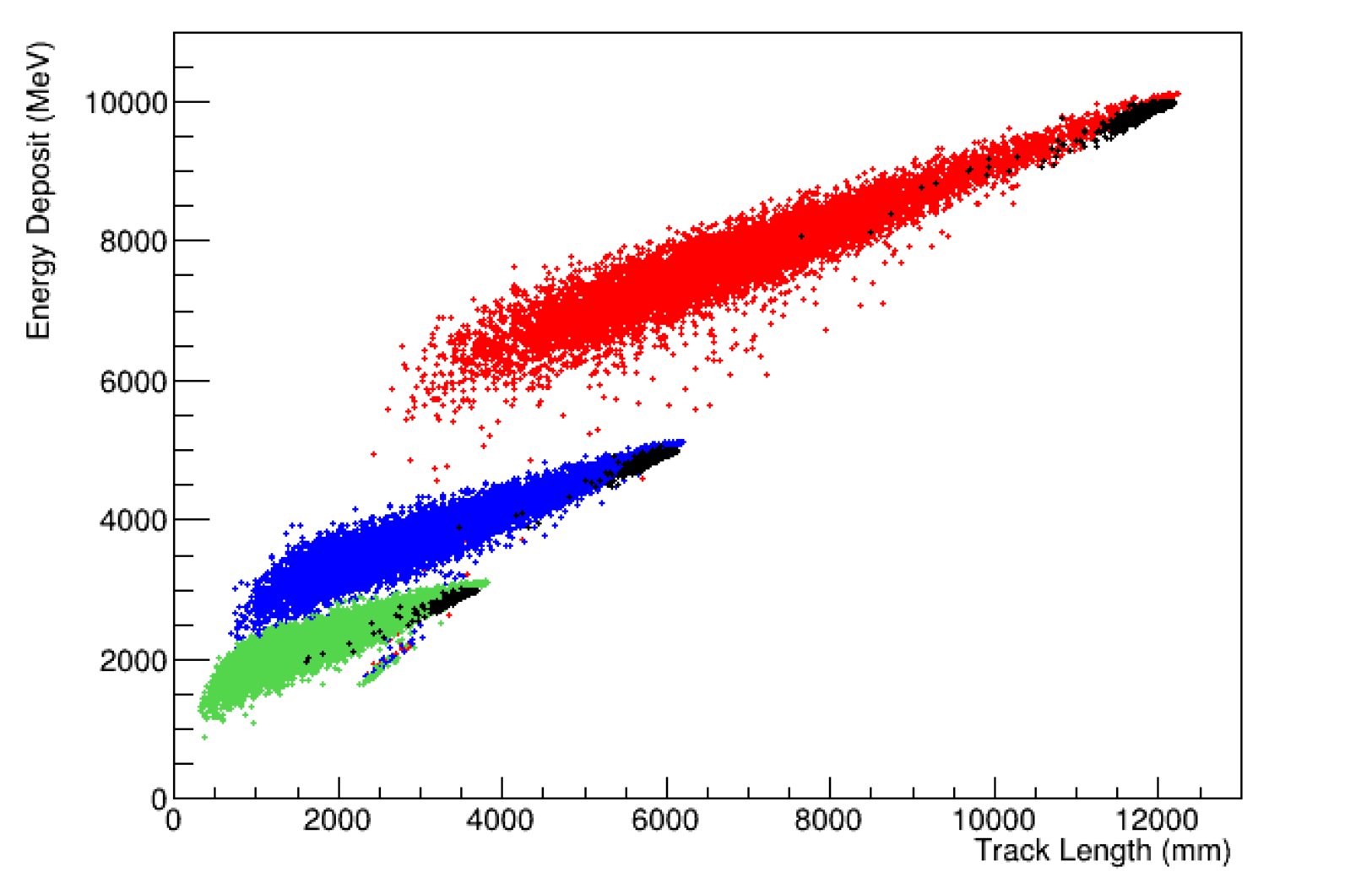}
  \caption{ Correlation plot between energy deposition and track length for a homogeneous glass scintillator calorimeter. The plot showcases the relationship for incident positive pions and electrons.  The pion data are displayed using various colors, with red indicating 10GeV, blue representing 5GeV, and green denoting 3 GeV). The electrons are presented by black dots and corresponded to three different energies (10GeV, 5GeV, and 3GeV).}
  \label{fig:1}
\end{figure}

We have previously established a correlation between energy measurements and track length of high-energy hadrons and electron showers \cite{bib:2}.
This relationship is depicted in Figure \ref{fig:1} as a scatter plot, where different energetic particles are presented by distinct colors.

The investigation yielded four significant findings.
\begin{enumerate}
\item A conspicuous and robust correlation between the two parameters; energy deposition in scintillation glasses and track length in lead glasses.
\item The correlation line, represented by a one-dimensional function, exhibits a consistent and unvarying slope across different energies.
\item The intercept of the linear fit line is directory proportional to the energy of the injected particle, thus indicating excellent linearity of the calorimeter.
\item The linear correlation pattern is observable across various particle types, including pions, kaons, protons, neutrons, and electrons.
\end{enumerate}

In Figure \ref{fig:1}, we also include the overlay of an electromagnetic (EM) shower events by electrons. The maximum energy observed in hadron showers corresponds to fully EM shower energy events, including of neutral pions. This region corresponds to the EM shower events of electrons, which are represented by black dots in the highest energy range of the figure.

To realize a large homogeneous calorimeter in practice, several challenges need to be addressed, including cost, light absorption, and precise separation of scintillation light from the relatively weak Cerenkov light.
In this paper, we present an innovative solution that offers a cost-effective, radiation-tolerant, and straightforward approach to overcome these challenges. 
Our proposed method involves utilizing a sandwich structure composing two distinct types of materials; a scintillator glass with the ability to emit scintillation photons, and an ordinary glass specifically chosen for Cerenkov light measurement. By incorporating these two materials in the calorimeter design, we can effectively separate the Cerenkov light from the scintillation photons with high precision.

The advantages of this approach are manifold.
Firstly, it offers an inexpensive alternative compared to homogeneous calorimeter, making it more accessible for large-scale calorimeter implementations.
Additionally, the proposed solution demonstrates excellent radiation tolerance, ensuring the long-term stability and durability of the calorimeter under high-energy particle interactions.

To validate the effectiveness of our proposed approach, we conducted extensive simulations to evaluate its performance. The simulation results provide compelling evidence of the method's capabilities in separating and accurately measuring both scintillation light and Cerenkov light within the calorimeter structure.

\section{Simulation and results}

The simulation conducted for this study assumed a simplified configuration involving a combination of glass-scintillator and lead glass layers.
However, it is important to note that in an actual detector, each layer would consists of 3cm$\times$3cm$\times$1cm thick tiles.
The calorimeter system would comprise 50 layers, stacked together to form a three-dimensional structure.
This design follows the guidelines of PFA directed calorimeters, such as the one utilized in the International Large Detector (ILD) \cite{bib:3}. 

To facilitate the readout of both scintillation and Cerenkov light, a silicon photo-sensor on tile technology allows for efficient detection and measurement of both types of light within the calorimeter system \cite{bib:4}.

\begin{figure}[ht!]
  \centering
   \includegraphics[width=80mm]{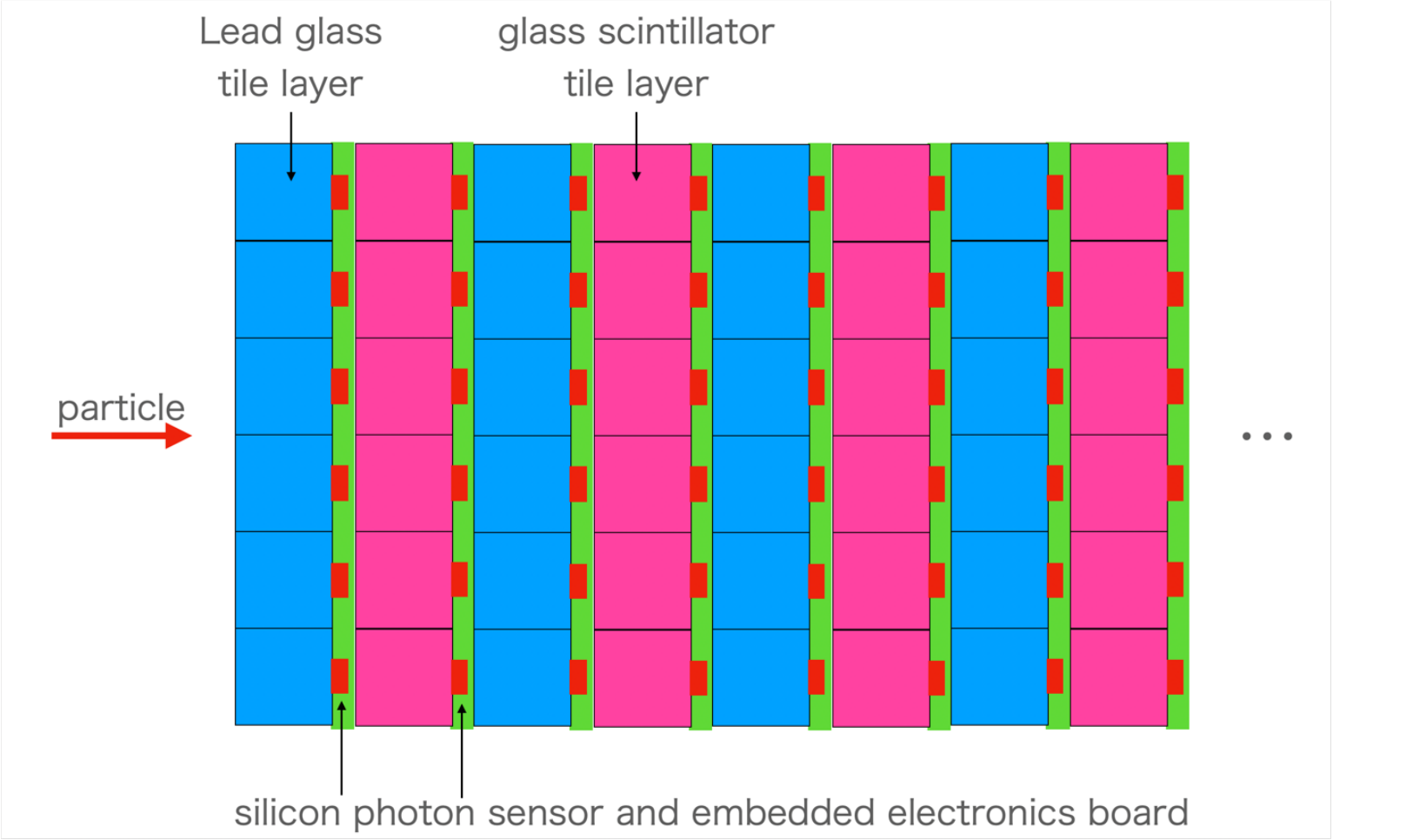}
  \caption{ Schematic of the proposed double readout sandwich calorimeter. The calorimeter structure comprises two distinct layers; lead glass and scintillator glass. Each layer consists of the tiles, meticulously designed to meet PFA. To facilitate efficient readout and measurement, each tile is equipped with a silicon photo-sensor. The photo-sensor captures and detects both scintillation and Cerenkov light emitted by the tile. The electrical signals generated by the photo-sensor are processed on the a dedicated readout board within the layer, directly connected to the tile. Subsequently, these processed signals are transmitted to the external Data Acquisition (DAQ) system for further analysis and processing.}
  \label{fig:2}
\end{figure}

The combination of lead glass and scintillation calorimeters, forming a Double readout Sandwich Calorimeter (DSC) (as depicted in Figure \ref{fig:2}), ensures optimal performance and enables the accurate determination of energy deposit.

The DSC configuration, with its innovative readout system, offers several advantages.
It enhances the precision and resolution of energy measurements, allowing for finer granularity and improved spatial resolution within the calorimeter.
This is crucial for capturing and analyzing the intricate details of particle showers and accurately reconstructing their energy profiles.

\begin{figure}[ht!]
  \centering
   \includegraphics[width=80mm]{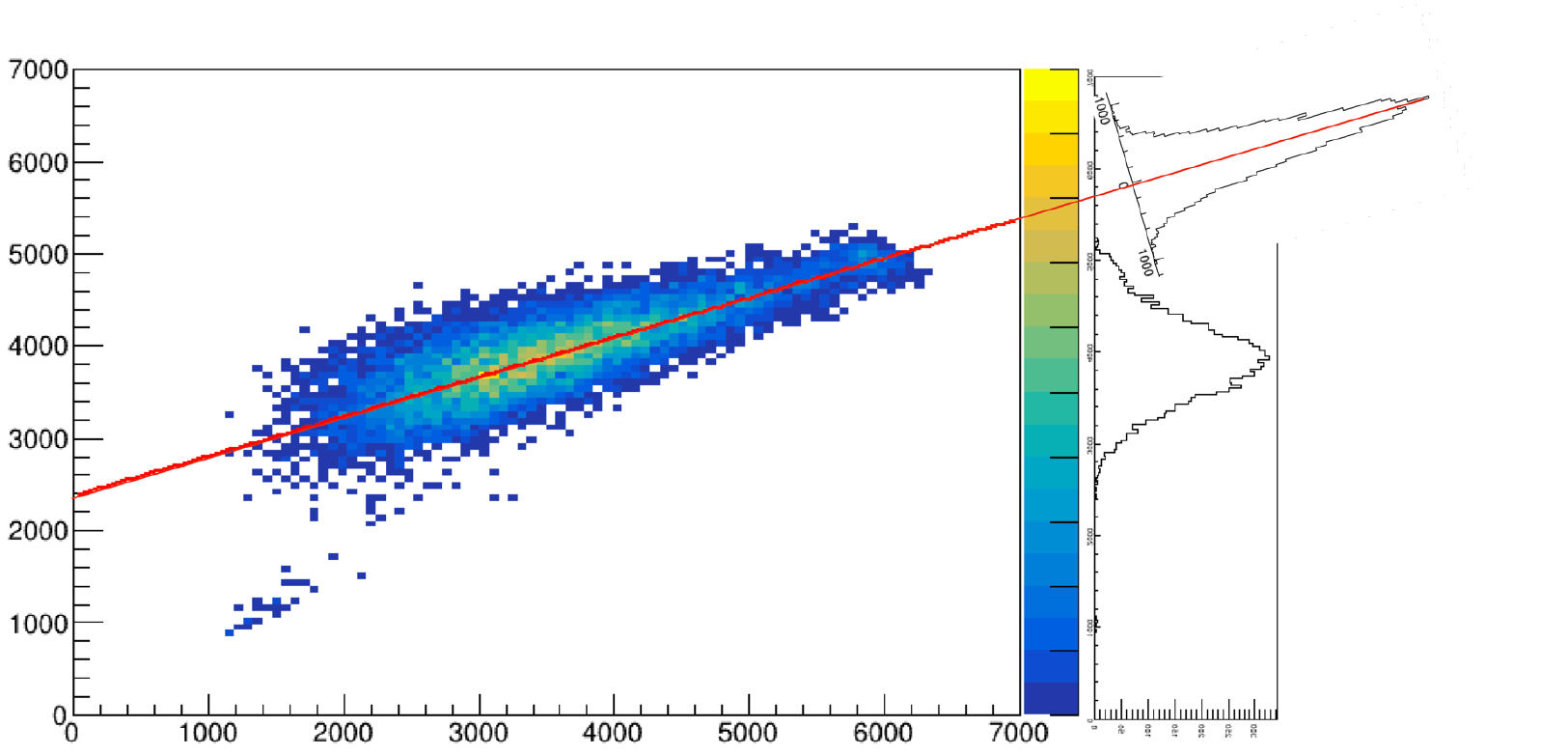}
  \caption{ A correlation plot of the energy deposit (ED) versus and track length (TL) for 10GeV pions 10,000 events. Notably, a clear correlation is observed. Additionally, projected plots are included in this view, specifically representing the energy sum measured in the scintillation glass layers only. Moreover, the optimized energy resolution of the calorimeter, determined by calculating the shortest distance to the one-dimensional fitting line in the correlation plot, is plotted alongside the correlation plot. }
  \label{fig:3}
\end{figure}

In the simulation conducted to assess the physics performances of the calorimeter, we simplified the model by neglecting the effects of silicon photo-sensors and embedded electronics layers. This simplification allowed us to focus specifically on the calorimeter response.
Consequently, the energy deposition was measured solely in the scintillation glass layers, while the track length was measured exclusively in the lead glass layers for positive pions of 10GeV. The correlation between energy deposition and track length is presented in Figure \ref{fig:3}, resembling the correlation plot observed for the homogeneous calorimeter (Figure \ref{fig:1}). 

The overall dimensions of the simulated calorimeter system were set to 2m$\times$2m$\times$2m, while the tile sizes standardized at 3cm$\times$3cm$\times$1cm thick for both scintillator and Cerenkov layers.
For consistency, we employed identical lead glasses in both the scintillation and Cerenkov layers, ensuring that they had the same components and density, which was measured to be 6.220g/cm3.

To conduct the simulations, we utilized the GEANT4 10.07 simulation framework, employing default parameters. The hadron interaction model employed is FTFP\_BERT \cite{bib:5-1}\cite{bib:5-2}, which provides a reliable representation of hadronic interactions within the simulated environment. By adopting this simulation framework and incorporating these specific parameters, we aimed to accurately assess the performance of the proposed calorimeter design and evaluate its effectiveness in measuring energy deposition and track length for various particles and energy ranges.

The correlation plot showcases the relationship between energy deposit and track length, revealing the distinctive behavior of different particles.
Among them, 10 GeV positive pions exhibit a particularly pronounced correlation pattern.
Furthermore, to provide additional insight, projected plots are included to highlight the energy sum exclusively measured in the scintillation glass layer.
These projections offer a focused view of the energy deposition characteristics within the calorimeter.

Additionally, the optimized energy resolution of the calorimeter is determined through a meticulous analysis of the correlation plot. By calculating the shortest distance to the one-dimensional fitting line, the optimized energy resolution is obtained. This information is plotted alongside the correlation plot, providing valuable insights into the performance and capabilities of the proposed calorimeter design.

\begin{figure}[ht!]
  \centering
   \includegraphics[width=80mm]{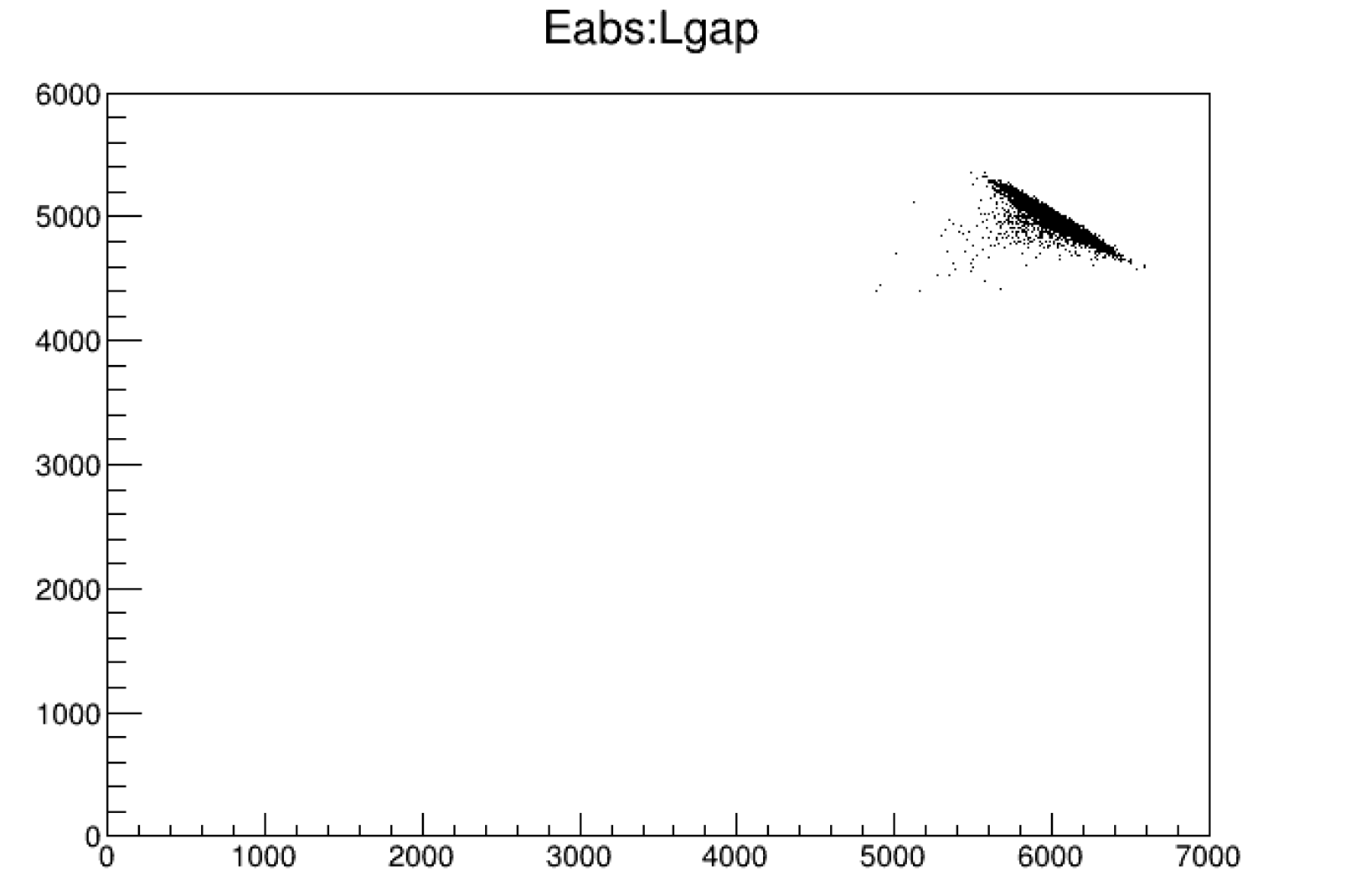}
  \caption{ Response of DSC for 10 GeV positrons $^+$. The vertical axis represents the response of scintillation glass layer, measured in MeV units. Meanwhile, the horizontal axis displays the response of the lead glass layers, represented as track length in millimeter (mm) units. }
  \label{fig:4}
\end{figure}

The plotted data in Figure \ref{fig:4} provide insights into the response characteristics of the DSC when subjected to 10 GeV positrons. The response of the scintillation glass layer, indicative of the energy deposited by the particles, is depicted on the vertical axis in MeV units. On the horizontal axis, the response of the lead glass layers is presented as track length, allowing for the measurement of the particle's trajectory within the calorimeter in millimeter units (mm).

By analyzing the response data in this configuration, it becomes possible to discern the behavior and performance of the DSC for 10 GeV positrons. The response characteristics of the scintillation and lead glass layers provide valuable information regarding the energy deposition and track length measurements, respectively, contributing to a comprehensive understanding of the calorimeter's capabilities and its suitability for high-energy particle detection. Since the both glass measure the electromagnetic (EM) shower, the sum of these two components remains constant, indicating a linea response characteristic for electrons.

The energy resolution of DSC is determined by projecting the line of linear fit to the hadronic clusters.
This projection distance necessitates a linear transformation and a normalized factor, denoted as 'c',  derived from the energy deposition (ED) and track length (TL) measurements. Specifically, the parameter 'c' can be obtained from the fitted straight line equation $ED = a \times TL + b $ as $ c = 1/ \sqrt{1 + a^2 }$, where 'a' and 'b' represent the linear fitting parameters as usual.
By applying this projection and normalization, the energy resolution can be determined at the incident beam energy.
The projected energy distribution provides valuable insights into the DSC's energy resolution performance.
Table \ref{table:1} presents a comparison of the energy resolutions achieved by the DSC, the glass scintillator homogeneous calorimeter, and the conventional energy deposition sum calorimeter of glass scintillator. The specific energy resolution values for each type of calorimeter are listed, highlighting the performance of the DSC relative to the other calorimeter configurations.

\begin{table}[h!]
\centering
\begin{tabular}{ |p{2.5cm}|p{2.5cm}|p{2.5cm}|p{2.5cm}| }
 \hline
 Positive pion energy (GeV) & Energy resolution (\%) for DSC & Energy resolution (\%) for homogeneous calorimeter & Energy resolution (\%) for energy deposit calorimeter \\ 
 \hline
 2 & 5.11 & 4.88 & 14.18 \\ 
 3 & 4.99 & 4.11 & 12.68 \\  
 4 & 4.35 & 3.59 & 12.16 \\  
 5 & 3.94 & 3.14 & 11.58 \\  
 10 & 2.46 & 2.10 & 8.76 \\  
 20 & 1.71 & 1.63 & 6.78 \\  
 30 & 1.40 & 1.57 & 6.12 \\  
 50 & 1.21 & 1.51 & 5.54 \\  
 100 & 1.10 & 1.65 & 5.07 \\
 \hline
\end{tabular}
\caption{
Table of the energy resolution of three different glass calorimeters as a function of incident positive pion energy. (a) glass scintillator and lead glass sandwich calorimeter (first row), (b) scintillator glass homogeneous calorimeter (second row), and (c) traditional energy measuring scintillation glass calorimeter (third row).
}
\label{table:1}
\end{table}

In this study, we compared the performance of the proposed DSC with that of a homogeneous calorimeter and a conventional calorimeter. 

A homogeneous calorimeter measures the total energy deposit of the glass scintillator and lead glass layers and as well as the sum of the track lengths of both the scintillator and lead glass layers, followed by one-dimensional fitting to create a correlation line and projected to be perpendicular to the line. The energy resolution is determined by projecting the data perpendicular to this correlation line. On the other hand, DCS calorimeter measures the energy information from the scintillation glass layer, and measures the track length from the lead glass layer separately, and creates a correlation. In contrast, a conventional calorimeter consists of a fully active scintillator glass layer and measures the total energy deposit within the calorimeter without distinguishing between different materials or layers.

\begin{figure}[ht!]
  \centering
   \includegraphics[width=80mm]{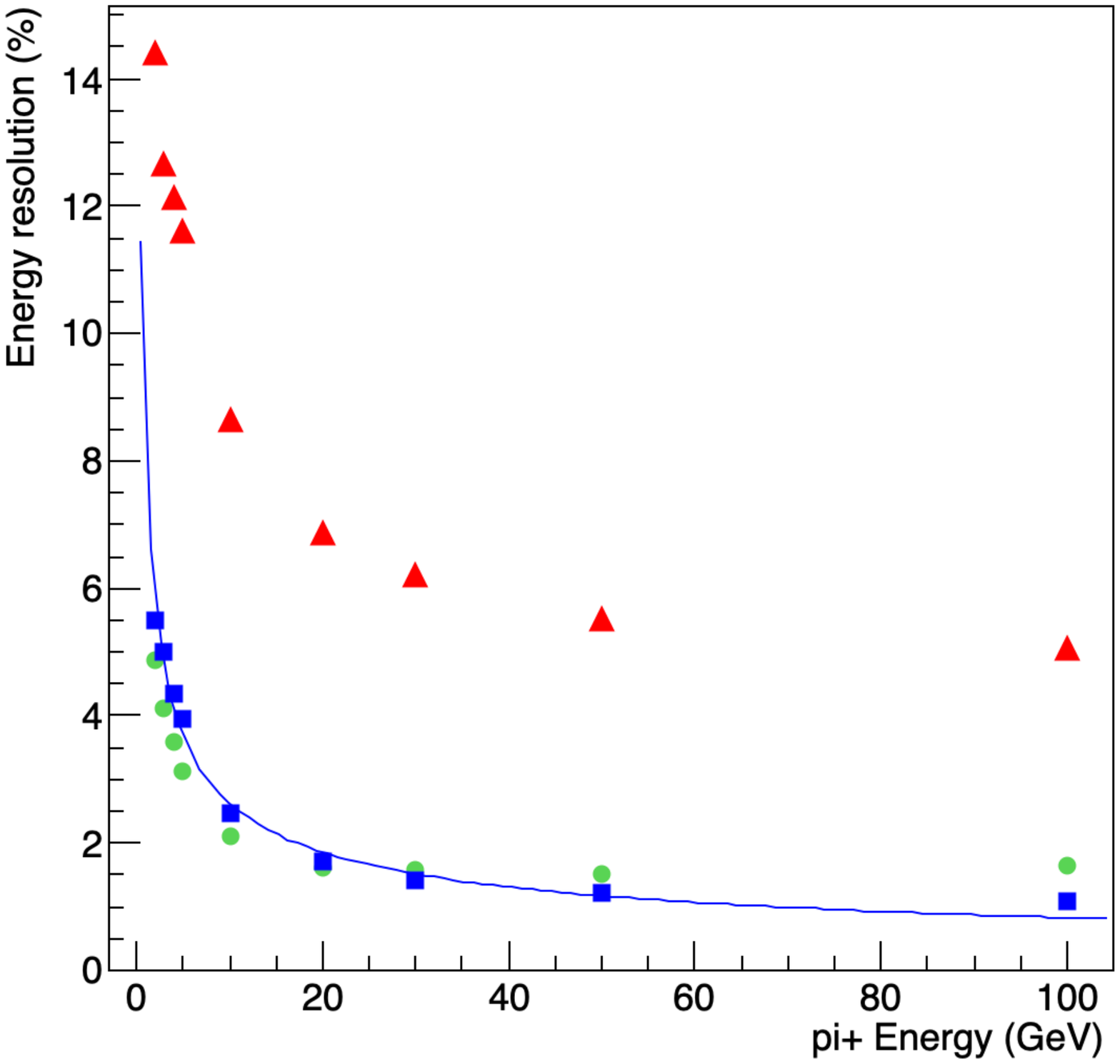}
  \caption{ Eenrgy resolution of hadron calorimeters for high energy positive pios. Three different types of calorimeters are compared : conventional calorimeter (red triangles), homogeneous calorimeter (green circles)  and DSC calorimeter (blue squares). The blue line represents the energy resolution of $ 9\% / \sqrt{E(\textrm{GeV})}$. }
  \label{fig:5}
\end{figure}

Figure \ref{fig:5} presents a comparison of the energy resolution performance of these three types of calorimeters as a function of particle energies.

By evaluating the performance of these calorimeters, we can assess their effectiveness in accurately measuring and characterizing energy deposition and track length of particles. The DSC calorimeter offers the advantage of separate measurement and correlations in a straightforward manner, allowing for a more detailed analysis of particle interactions within the calorimeter, especially when the layers are subdivided into tiles. The DSC calorimeter demonstrates an impressive energy resolution of approximately $ 9\% / \sqrt{E(\textrm{GeV})}$, which is exceptional for a high-energy hadron calorimeter. This remarkable energy resolution positions the DSC calorimeter as a critical component in future calorimeter designs.

One of the key strengths of the DSC calorimeter lies in its finely segmented three-dimensional structure, allowing for precise localization and measurement of energy deposits.
This segmentation enhances the calorimeter's ability to accurately capture and analyze particle interactions.

Additionally, the DSC calorimeter offers excellent timing capabilities due to its Cerenkov photo-detection mechanism.
This timing precision provides valuable information about the particle's trajectory and enables advanced event reconstruction techniques.

As a result of these remarkable features, DSC is expected to play a vital role in the development of future calorimeters.
Their finely segmented structure, exceptional energy resolution, and enhanced timing capabilities make them an appealing choice for high-energy physics experiments, enabling more accurate and detailed measurements of hadron showers.

In the simulations conducted for this study, the effect of photon statistics was not considered.
However, it is important to note that the small size of the tiles in the calorimeter can significantly improve the energy measurement by ensuring a large number of tiles for the smallest energy deposit caused by ionizing particles.
This abundance of measurements contributes to a more accurate determination of the energy deposited by particles. 

 Another aspect to consider is the reabsorption and conversion of Cerenkov light into scintillation light within the glass scintillator.
The light absorption properties of the scintillation glass itself offer the potential to minimize the contribution of Cerenkov light to the scintillation light. Typically, the glass exhibits strong absorption at wavelengths of 400 nm or less, effectively absorbing the majority of Cerenkov light generated within the scintillation glass.
 However, experimental validation is required to confirm this assumption.
 The development of heavy scintillator glasses that provide sufficient light yield for minimum ionization particles (MIP) is a significant challenge for future advancements.
 It is desirable to explore the possibility of developing glass scintillators with high density and increased light output.
 Quantum dots technology, which has gained attention in recent years, holds promise for achieving these objectives.
 Urgent efforts should be directed towards the development and implementation of such technologies.
 While ensuring the mechanical stability of the calorimeter structure is crucial, it may be possible to enhance stability by introducing a mega-tile structure that subdivides a large plated glass.
 This approach could potentially improve the overall mechanical integrity of the calorimeter system.
 
 In summary, future research endeavors should focus on addressing the impact of photon statistics, validating the reduction of Cerenkov light absorption, developing heavy scintillator glasses with high light yield, and exploring innovative technologies such as Quantum dots.
 Additionally, exploring structural improvements such as the mega-tile design could enhance the mechanical stability of the calorimeter.
 These advancements will contribute to the overall performance and feasibility of the proposed Double readout Sandwich Calorimeter.

\end{document}